\documentclass[twocolumn,secnumarabic,amssymb, nobibnotes, aps, pra]{revtex4-1}

\usepackage{graphicx, color, braket, amsmath}
\makeatletter
\renewcommand*{\@fnsymbol}[1]{\ifcase#1\or\else\fi}
\makeatother

\begin{document}

\title{Quantum optical neural networks}%

\begin{abstract}
Physically motivated quantum algorithms for specific near-term quantum hardware will likely be the next frontier in quantum information science. 
Here, we show how many of the features of neural networks for machine learning can naturally be mapped into the quantum optical domain by introducing the quantum optical neural network (QONN).
Through numerical simulation and analysis we train the QONN to perform a range of quantum information processing tasks, including newly developed protocols for quantum optical state compression, reinforcement learning, and black-box quantum simulation.
We consistently demonstrate our system can generalize from only a small set of training data onto states for which it has not been trained.
Our results indicate QONNs are a powerful design tool for quantum optical systems and, leveraging advances in integrated quantum photonics, a promising architecture for next generation quantum processors.
\end{abstract}

\author{Gregory R. Steinbrecher$^{1}$}
\author{Jonathan P. Olson$^{2}$}
\author{Dirk Englund$^{1}$}
\author{Jacques Carolan$^{1*}$}
\thanks{$^{*}$carolanj@mit.edu}

\affiliation{$^{1}$Research Laboratory of Electronics, Massachusetts Institute of Technology, Cambridge, MA, USA 02139}
\affiliation{$^{2}$Zapata Computing Inc., 501 Massachusetts Ave., Cambridge, MA, USA 02139}
\date{\today}%
\maketitle

\section{Introduction} % (fold)
\label{sec:introduction}

% section introduction (end)
Deep learning is revolutionizing computing \cite{LeCun:2015dt, Goodfellow:2016wc} for an ever-increasing range of applications, from natural language processing \cite{Wu:2016wt} to particle physics \cite{Radovic:2018iz} to cancer diagnosis \cite{Capper:2018dy}.
These advances have been made possible by a combination of algorithmic design \cite{Glorot:2011tm} and dedicated hardware development \cite{Sze:2017ka}. 
Quantum computing \cite{Nielsen:2011vx}, while more nascent, is experiencing a similar trajectory, with a rapidly closing gap between current hardware and the scale required for practical implementation of quantum algorithms.
Error rates on individual quantum bits (qubits) have steadily decreased \cite{Barends:2014fu, Harty:2014cm}, and the number and connectivity of qubits have improved \cite{Bernien:2017bp, Zhang:2017et}, making so-called Noisy Intermediate Scale Quantum (NISQ) processors \cite{Preskill:2018uv} capable of tasks too hard for a classical computer a near-term prospect. 
Experimental progress has been met with algorithmic advances \cite{Montanaro:2016iz} and near-term quantum algorithms have been developed to tackle problems in combinatorics \cite{Farhi:2014wk}, quantum chemistry \cite{AspuruGuzik:2012ho} and solid state physics \cite{Wecker:2015kd}.
However, it is only recently that the potential for quantum processors to accelerate machine learning has been explored \cite{Biamonte:2017ic}.

Quantum machine learning algorithms for universal quantum computers have been proposed \cite{Harrow:2009gx, Lloyd:2014gc, Rebentrost:2014fi} and small-scale demonstrations implemented \cite{Cai:2015jk}, though the requirements for practical protocols remain an open question \cite{Aaronson:2015hl}.
Relaxing the requirement of universality, quantum machine learning for NISQ processors has emerged as a rapidly advancing field \cite{Mitarai:2018tv, Farhi:2018wv, Schuld:2018vp, Havlicek:2018tu} that may provide a plausible route towards practical quantum-enhanced machine learning systems. 
These protocols typically map features of machine learning algorithms (such as hidden layers in a neural network) directly onto a shallow quantum circuits in a platform independent manner.
In contrast, the work presented here leverages features unique to a particular physical platform. 

In this work, we introduce an architecture for neural networks unique to quantum optical systems: the Quantum Optical Neural Network (QONN).
We argue that many of the features which are natural to quantum optics (mode mixing, optical nonlinearity) can directly be mapped to neural networks.
Moreover, technological advances driven by trends in photonic quantum computing \cite{OBrien:2007ioa, Obrien:2009eu, Rudolph:2017du} and the microelectronics industry \cite{Sun:2015gg} offer a plausible route towards large-scale, high-bandwidth QONNs, all within a CMOS compatible platform.

Through numerical simulation and analysis, we apply our architecture to a number of key quantum information science protocols.
We benchmark the QONN by designing quantum optical gates where circuit decompositions are already known.
Next, we show that our system can learn to simulate other quantum systems using only a limited set of input/output state pairs, generalizing what it learns to previously unseen inputs.
We demonstrate this learning on both Ising and Bose-Hubbard Hamiltonians. 
We then introduce and test a new quantum optical autoencoder protocol for data compression, with applications in quantum communication and quantum networks, which again relies on the ability to train our systems using a subset of possible inputs. 
Finally, we apply our system to a classical machine learning controls task, balancing an inverted pendulum by a reinforcement learning approach.
Our results may find application both as an important technique for designing next generation quantum optical systems, as well as a versatile experimental platform for near-term optical quantum information processing and machine learning.

\begin{figure*}[t!]
  \begin{centering}
    \includegraphics[width=7in]{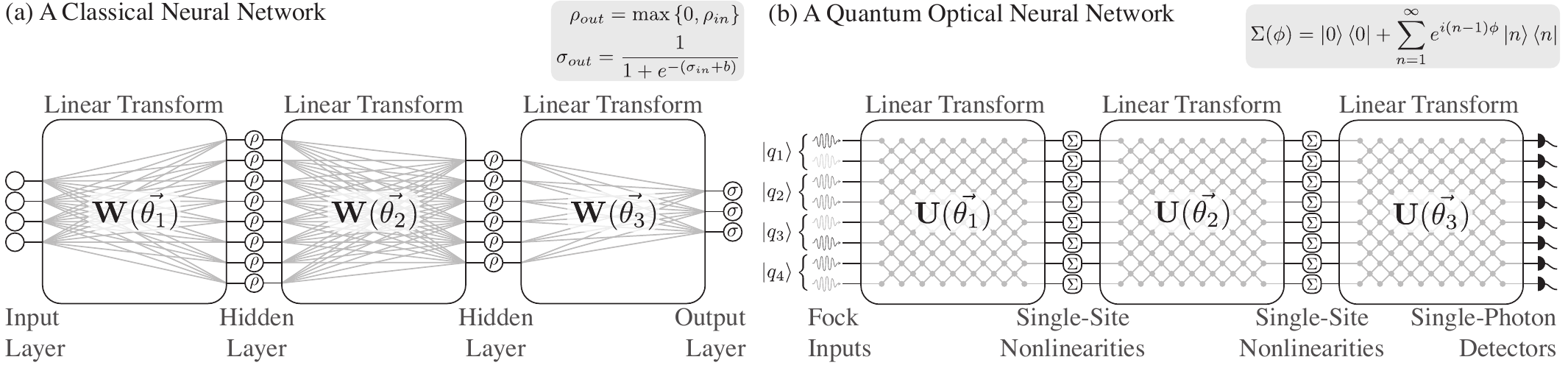}
  \end{centering}
  \caption{\label{fig:Architecture} 
\textbf{Quantum Optical Neural Network.} 
(a) An example of a classical neural network architecture. Hidden layers are rectified linear units (ReLU) and the output neuron uses a sigmoid activation function to map the output into the range $(0,1)$. 
(b) An example of our quantum optical neural network (QONN) architecture. Inputs are dual-rail Fock states which encoded qubits, with a photon in the top mode representing $\left|0\right>$ and a photon in the bottom mode representing $\left|1\right>$. 
The single-site nonlinearities are given by $\chi^{(3)}$ functions: a Kerr-type interaction applying a constant phase for each additional photon present. 
Readout is given by single photon detectors which measure the photon number at each output mode.}
\end{figure*}

In prototypical neural networks [see Fig.\ref{fig:Architecture}(a)] an input vector $\vec{x}\in \mathbb{R}^n$ is passed through multiple layers of: 
(1) linear transformation, i.e.\ a matrix multiplication $W(\theta_i).\vec{x}$ parameterized by weights $\theta_i$ at layer $i$, and 
(2) nonlinear operations $\sigma(\vec{x})$ which are single site nonlinear functions sometimes parameterized by biases $\vec{b}_{i}$ (typically referred to as the perceptron or neuron, see Fig.\ref{fig:Architecture}(a), inset for two examples: the rectifying neuron and the sigmoid neuron).
The goal of the neural network is to optimize the parameter sets $\{\theta_i\}$ and $\{b_i\}$ to realize a particular input-output function $f(\vec{x})=y$.
The power of neural networks lies in the fact that when trained over a large data set $\{\vec{x}_i\}$, this often highly nonlinear functional relationship is generalizable to a large vector set to which the network was not exposed during training.
For example, in the context of cancer diagnosis, the input vectors may be gray scale values of pixels of an image of a cell, and the output may be a two dimensional vector that corresponds to the binary label of the cell as either a benign or malignant \cite{Kourou:2015jx}.
Once the network is trained, it may categorize with high probability new, unlabelled, images of cells as either `benign' or `malignant'.  

A number of the key components of classical neural networks are readily implementable using state of the art integrated quantum photonics.
First, matrix multiplication can be realized across optical modes (where each mode contains a complex electric field component) via arrays of beamsplitters and programmable phase shifts \cite{Reck:1994dz, Clements:2016tv}.
In the lossless case, an $n$-mode optical circuit comprising $n(n-1)$ components implements an arbitrary $n\times n$ single particle unitary operation (which can also be used for classical neural networks \cite{Arjovsky:2015tb, Jing:2016tk}), and a $n$-dimensional non-unitary operation can always be embedded across a $2n$-mode optical circuit \cite{Miller:2013ij}.
Advances in integrated optics have enabled the implementation of such circuits for applications in quantum computation \cite{Carolan:2015vga}, quantum simulation \cite{Lanyon:2009jf, Sparrow:2018ba}, and classical optical neural networks \cite{Shen:2017hb}.
Second, optical nonlinearities are a core component of many classical \cite{Miller:2009fi, Miller:2010bm} and quantum \cite{Knill:2001vi, Kok:2007ep} optical computing architectures.
Single photon coherent nonlinearities can be implemented via measurement \cite{Knill:2001vi}, interaction with three-level atoms \cite{Duan:2004gg} or superconducting materials \cite{Kirchmair:2013gu}, and through all-optical phenomena such as the Kerr effect \cite{Brod:2016ji}.
While integration of each of these technologies into a single scalable system is an outstanding challenge for the field, for generality, the architecture we present considers idealized components.
In this work we focus on discrete variable QONNs due to the maturity of the technology platform, but note that continuous variable implementations are also promising \cite{killoran2018continuous}.

\begin{figure*}[t!]
  \begin{centering}
    \includegraphics[width=7in]{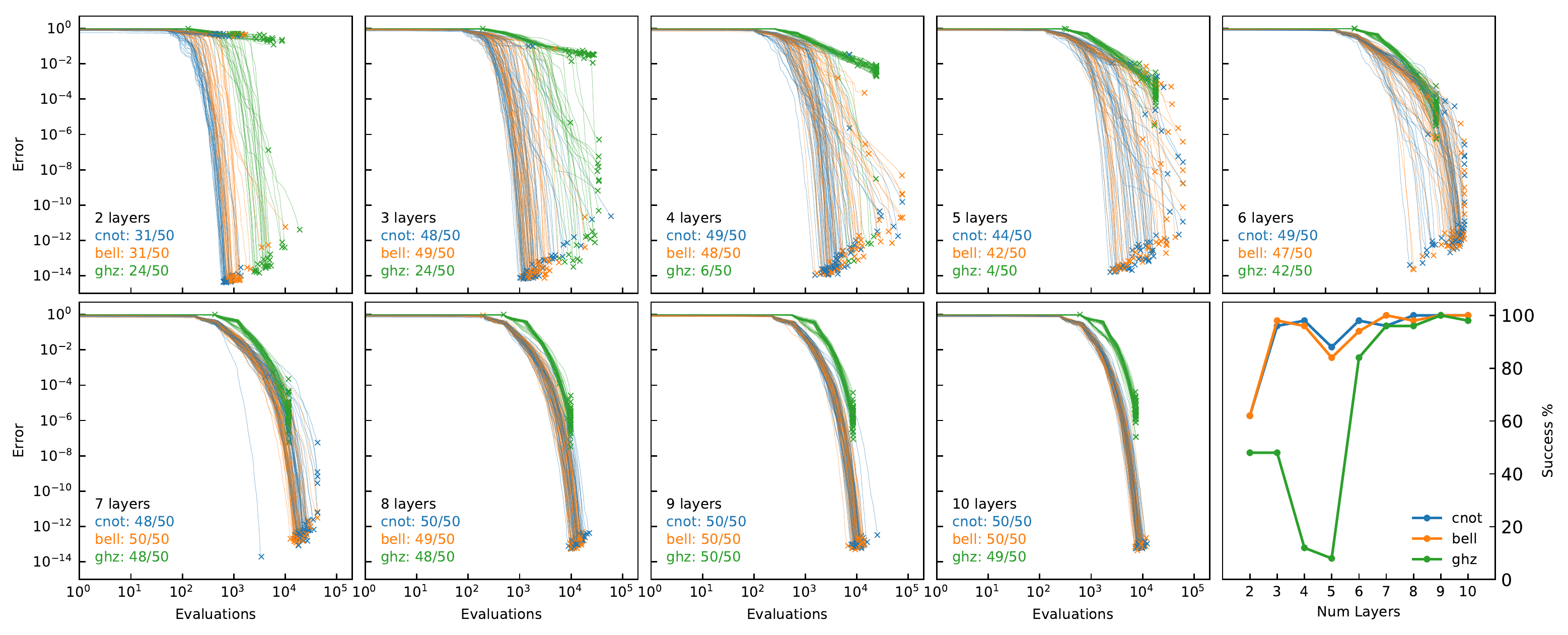}
  \end{centering}
  \caption{\label{fig:Benchmarking} 
\textbf{Benchmarking Results.} The first nine figures show 50 training runs for each of three representative optical quantum computing tasks: performing a CNOT gate, separating/generating Bell states, and generating GHZ states. At low layer depth, the optimizations frequently fail to converge to an optimal value (we defined an error less than $10^{-4}$ as ``success''), terminating at relatively large errors. This behavior gets worse as we add layers, out to 5 layers, at which point it undergoes a rapid reversal, with the training essentially always succeeding at layer depths of 7 or more. This is shown in the final figure, where success percentage is plotted against the number of layers for each of the three tasks. 
}
\end{figure*}

\section{Architecture} % (fold)
\label{sub:mathematical_formalism}

As shown in Fig.~\ref{fig:Architecture}(b), input data to our QONN is encoded as photonic quantum states $\ket{\psi_i}$, either as dual rail qubits requiring two optical modes per photon ($\ket{0}\equiv\ket{10}_{12}, \ket{1}\equiv\ket{01}_{12}$), or more generally as a Fock states $\ket{i}_j$ (corresponding to $i$ photons in the $j^\text{th}$ optical mode), which for $n$ photons in $m$ modes is described by a $\binom{n+m-1}{m}$-dimensional complex vector of unit magnitude.
The linear circuit is described by an $m$-mode linear optical unitary $U(\vec{\theta})$ parameterized by a vector $\vec{\theta}$ of $m(m-1)$ phases shifts $\theta_i \in (0,2\pi]$ via the encoding of Reck et al.,\ \cite{Reck:1994dz}.
The nonlinear layer $\Sigma$ comprises single mode $\chi^{(3)}$ interactions in the monochromatic approximation, applying a constant phase for each additional photon present via self-phase modulation \cite{Brod:2016ji}. For a given interaction strength $\phi$, this unitary can be expressed as $\Sigma\left(\phi\right) =  \sum_{n=1}^{\infty} \left|0\right> \left<0\right| + e^{i(n-1)\phi} \left|n\right> \left<n\right|$.
The full system comprising $N$ layers is therefore
\begin{equation}\label{eq1}
	S(\vec{\Theta}) = \prod_i^N \Sigma(\phi) . U(\vec{\theta}_i),
\end{equation}
where $\vec{\Theta}$ is a $Nm(m-1)$-dimensional vector and the strength of the nonlinearity is typically fixed as $\phi=\pi$. 
Finally, single photon detectors will be used to measure the photon number at each output. We use the results of this measurement, along with a training set of $K$ desired input/output pairs $\left\{\ket{\psi^i_\text{in}}\rightarrow\ket{\psi^i_\text{out}}\right\}_{i=1}^K$, to construct a cost function
\begin{equation}
	C\left(\vec{\Theta}\right) = 1-1/K\sum_{i=1}^K |\braket{\psi^i_\text{out}|S(\vec{\Theta})|\psi^i_\text{in}}|^2
\end{equation}
that is variationally minimized over $\vec{\Theta}$.
% subsection mathematical_formalism (end)

We distinguish between two approaches to training: in situ and in silico.
The in situ approach directly optimizes the quantum optical processor and 
measurements are made via single photon detectors at the end of the circuit. 
One aim is to optimize figures of merit that can be estimated with a number of measurements that scales polynomially with the photon number (as opposed to full quantum process tomography \cite{OBrien:2004cn}). 
If the target state is accessible the overlap can be estimated with the addition of a controlled-SWAP operation, which is related to the Hong-Ou-Mandel effect in quantum optics \cite{GarciaEscartin:2013ie}. 
Efficient fidelity proxies provide another route towards estimating salient features of quantum states without reconstruction of the full density matrix \cite{Cramer:2010bs}.
Moreover, the in situ approach may enable a form of error mitigation by routing quantum information around faulty hardware \cite{Mower:2015co}.
In contrast, the in silico approach simulates the QONN on a digital classical computer and keeps track of the full quantum state internal to the system.
Simulations will therefore be limited in scale, but may help guide the design of, say, quantum gates where the optimal decomposition is not already known, or as an ansatz for the in situ approach.
In Appendix~\ref{sub:training}, we describe the computational techniques used in this work.

\section{Benchmarking}
As a first step in validating our architecture, we ensure it can learn elementary quantum tasks such as quantum state preparation, measurement and quantum gates. 
We chose Bell state projection/generation, GHZ state generation, and the implementation of the CNOT gate as representative of typical optical quantum information tasks.
As described in Appendix~\ref{sub:be}, in each of these cases the training set represents the full basis set for the quantum operation of interest, and successful training tells us something about the expressivity of our architecture.

We trained QONNs of increasing layer depth from $N=2\rightarrow 10$ with $\phi=\pi$. 
As shown in Fig.~\ref{fig:Benchmarking}, at short layer depths the optimization frequently terminates early, finding a non-optimal local minima. 
We observe similar behavior for all of the studied tasks.
Most notable here is the behavior of the optimization as the layer count increases: Just like a classical neural network, as we increase the layer depth, it becomes consistently easy to find a local minimum that is close to the global minimum.
This demonstrates the utility of deep networks: while a single layer may be \emph{sufficient} to implement a CNOT gate, with deep networks we can reliably discover a configuration that yields the correct operation.
For more complex operations, where the small-layer-number implementation may be difficult to find or simply not exist, this gives hope that we can still reliably train a deep network to perform the task.

\section{Hamiltonian Simulation} % (fold)
\label{sub:hamiltonian_learning}

\begin{figure}[t!]
\includegraphics[trim=0 0 0 0, clip, width=1.0\linewidth]{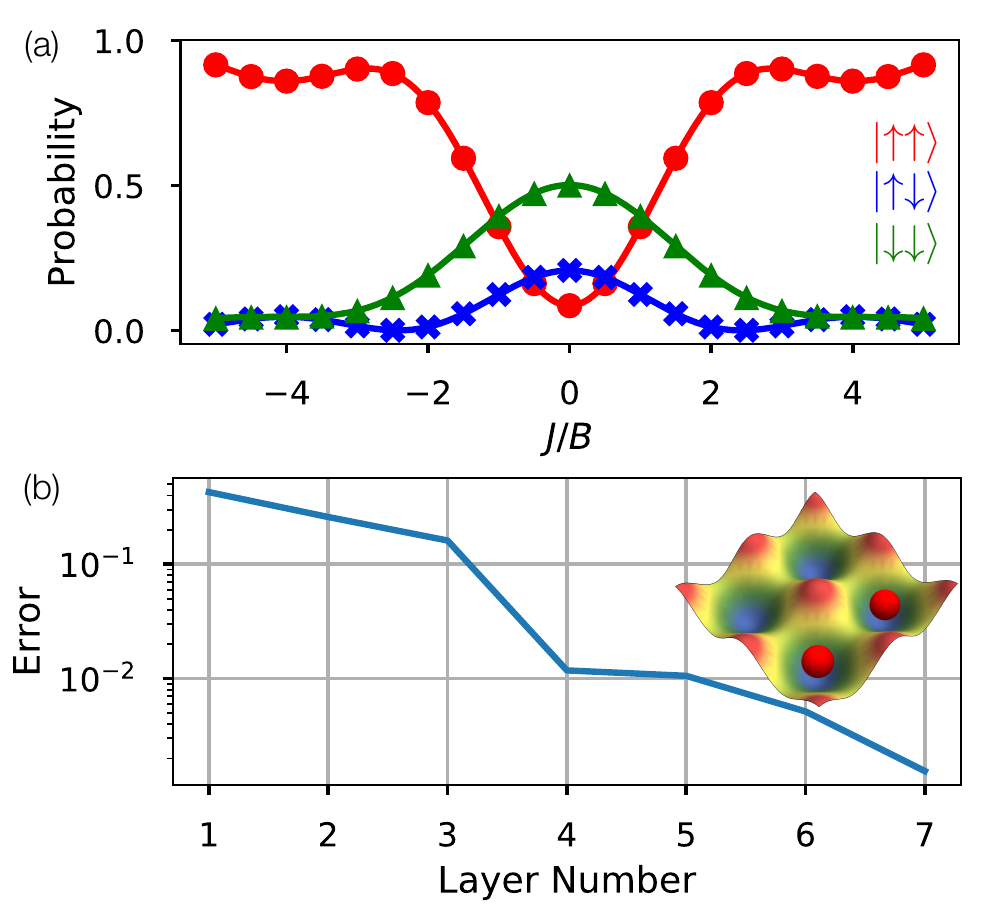}
\caption{\textbf{QONNs for Hamiltonian Simulation.}
(a) Ising Model. A three layer QONN is trained for a range of interaction strengths $J/B$ and the probability for particular output spin configuration is plotted (points) given the $\ket{\uparrow \uparrow}$ initialization state.
The expected evolution is plotted alongside (lines).
Critically, during the training process our QONN was never exposed to the initialization state.
(b) Bose-Hubbard model. Number of layers required to reach a particular test error for the simulation of a $(2,4)$ strongly interacting $U/t_\text{hop}=20$ Bose-Hubbard Hamiltonian (schematic shown in inset).  Training is performed 20 times for each layer depth, and the lowest test error is recorded.}
\label{fig:fig_sim}
\vspace{-0.0cm}
\end{figure}

While the results thus far benchmark the training of the QONN, a critical feature of any learning system is that it can generalize to states on which it has not been trained.
To assess generalization, we apply the QONN to the task of quantum simulation, whereby a well controlled system in the laboratory $S(\vec{\Theta})$ is programmed over parameters $\vec{\Theta}$ to mimic the evolution of a quantum system of interest described by the Hamiltonian $\hat{H}$.
In particular, we train our QONN on $K$ sets of input/output states  $\{\ket{\psi_\text{in}^i}\}$  $\{\ket{\psi_\text{out}^i}\}$ related by the Hamiltonian of interest $\ket{\psi_\text{out}^i}=\exp(-i\hat{H}t) \ket{\psi_\text{in}^i}$, and test it on new states which it has not been exposed to.

As a first test we look at the Ising model (see Appendix~\ref{sub:simulated_hamiltonians}) which is optically implemented via a dual-rail encoding with $m=2n$, where $\ket{\uparrow}\equiv\ket{10}_{12}$ and $\ket{\downarrow}\equiv\ket{01}_{12}$.
For the $n=2$ spin case, we train the QONN on a training set of 20 random two-photon states and test it on 50 different states.
We empirically determine that for a wide range of $J/B$ values (with $t=1$) a three layer QONN reliably converges to an optimum.
In Fig.~\ref{fig:fig_sim}(a) we vary the interaction strength $J/B\in [-5,5]$ and plot the probability of finding a particular spin configuration given an initialization state $\ket{\uparrow \uparrow}$.
Critically, this input state is not in set of states for which the QONN was trained.
We also train our QONN for the $n=3$ spin case, reaching an average test error of $ 10.1\%$.
This higher error in the larger system motivates the need for advanced training methods such as backpropagation \cite{Verdon:2018uw} or layer-wise training approaches \cite{Bengio:vb, Hettinger:2017wg} to efficiently train deeper QONN.

Finally we look at a Hamiltonian more natural for photons in optical modes, the Bose-Hubbard model, see Appendix~\ref{sub:simulated_hamiltonians} for further details.
Now, the $(n,m)$ configuration of bosons to be simulated is naturally mapped to an $n$-photon $m$-mode photonic system.

To benchmark our system we look at the number of layers required to express a $(2,4)$ strongly interacting Bose-Hubbard model with $U/t_\text{hop}=20$ and time parameter $t=1$.  
We constrain the connectivity to be that of a square lattice, as shown in Fig.~\ref{fig:fig_sim}(b), inset.
Figure.~\ref{fig:fig_sim}(b) shows that the linear (i.e. single-layer) system gives a mean error in the test set of $42\%$ and increasing the layer number steadily reduces this error to $0.1 \%$ at seven layers.
This suggests that deeper networks can express a richer class of quantum functions (i.e. Hamiltonians), a concept familiar in classical deep neural networks \cite{Raghu:2016wn}. 
Choosing five layers to give a reasonable trade-off between error ($\sim 1\%$) and computational tractability, we vary the interaction strength in the the range $U/t_\text{hop} \in [-20,20]$.
Across all experiments we achieve a mean test error of $2.9 \pm  1.3 \%$ (error given by the standard deviation in 22 experiments).

While our analysis has focused on Hamiltonians that exist in nature, the approach itself is very general: mimicking input-output configurations given access to a reduced set of input-output pairs from some family of quantum states.
This may find application in learning representations of quantum systems where circuit decompositions are unknown, or finding compiled implementations of known circuits.  

\section{Quantum Optical Autoencoder} % (fold)
\label{sub:quantum_optical_autoencoder}

Photons play a critical role in virtually all quantum communication and quantum networking protocols, either as information carriers themselves or to mediate interactions between long lived atomic memories \cite{Gisin:2007by}.
However, such schemes are exponentially sensitive to loss: given a channel transmissivity $\eta$ and number of photons $n$ required to encode a message, the probability of successful transmission scales as $\eta^n$.
Reducing the photon number while maintaining the information content therefore exponentially increases the communication rate.
In the following we use the QONN as a quantum autoencoder to learn a compressed representation of quantum states.
This compressed representation could be used, for example, to more efficiently and reliably exchange information between physically separated quantum nodes \cite{Kimble:2008if}.

Quantum autoencoders have been proposed as a general technique for encoding, or compressing, a family of states on $n$ qubits to a lower dimensional $k$-qubit manifold called the latent space \cite{Romero2017}.
Similar to classical autoencoders, a quantum autoencoder learns to generalize from a small training set $T$ and is able to compress states from the family that it has not previously seen.  
As well as applications in quantum communication and quantum memory, it has recently been proposed as a subroutine to augment variational algorithms in finding more efficient device-specific ansatzes \cite{Olson2018}.
In contrast, the quantum optical autoencoder encodes input states in the Fock basis.
Moreover, even if optical input states are encoded in the dual-rail qubit basis, the autoencoder may learn a compression onto a non-computational Fock basis latent space.

As a choice of a family of states, and one which is relevant to quantum chemistry on NISQ processors, we consider the set of ground states of molecular hydrogen, H$_2$, in the STO-3G minimal basis set \cite{Helgaker2013}, mapped from their fermionic representation into qubits via the Jordan-Wigner transformation \cite{Tranter.115.1431.IJQC.2015}. 
Ground states in this qubit basis (which we will denote with a subscript as the logical basis $L$) have the form $\ket{\psi(i)}=\alpha(i)\ket{0011}_L+\beta(i)\ket{1100}_L$, where $i$ is the bond length of the ground state. The qubits themselves are represented in a dual-rail encoding thus the network consists of $n=4$ photons in $m=8$ optical modes.

The goal of the quantum optical autoencoder $S$, is for all states in the training set $\ket{\psi_i}\in T$, satisfy
\begin{equation}
S\ket{\psi_i}=\ket{000}_L\ket{\psi_i^C},
\end{equation}
for some two-mode state $\ket{\psi_i^C}$ in the latent space.  
The quantum autoencoder can therefore be seen as an algorithm that systematically disentangles $n-k$ qubits from the set of input states and sets them to a fixed reference state (e.g. $\ket{0}_L^{\otimes n-k}$). 
For this reason, the fidelity of the reference state will be used a proxy for the fidelity of the decoded state. 

\begin{figure}[t!]
\centering
\includegraphics[width=1.0\linewidth]{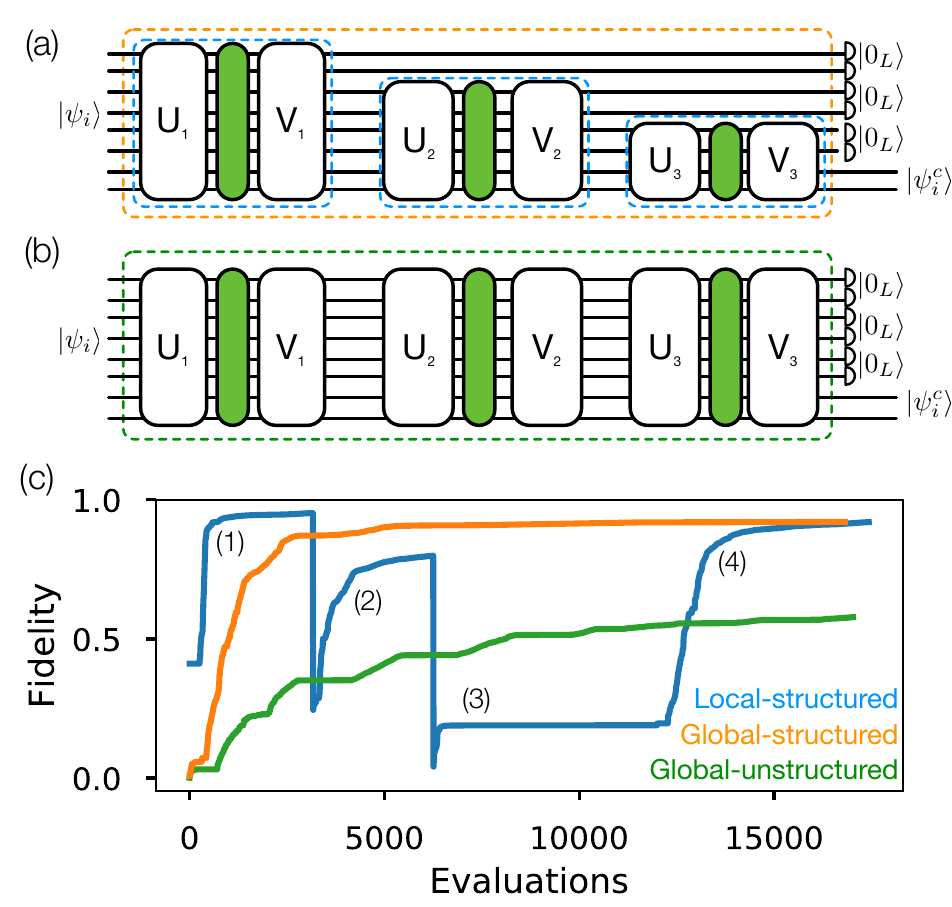}
\caption{ \textbf{Quantum Optical Autoencoder.} 
(a, b) Schematics of the QONN architectures corresponding to each of the three training strategies.  
While the architecture of the global-structured (a, orange) and global-unstructured (b, green) optimizations remained the same throughout the entire optimization, the local-structured approach (a, blue) optimized the parameters of (1) $U_1$ and $V_1$ first (with the nonlinear layer shown in green), before moving on to (2) $U_2$ and $V_2$, and in the third phase (3) $U_3$ and $V_3$.  The final refinement step of the iterative approach (4) considered all parameters in the optimization, similar to the global strategy.
(b) A plot of the fidelities of the reference states achieved by the different training strategies to compress ground states of molecular hydrogen.  While the global (orange) and unstructured (orange) optimizations included all three reference qubits from the start, the large drops in fidelity for the iterative procedure (blue) are due to including increasingly more reference states in the optimization.}
\label{fig:auto_encoder}
\end{figure}

To train a quantum autoencoder one should choose a circuit architecture with general enough operations to compress the input states, but few enough parameters to train the network efficiently.  
As shown in Fig.~\ref{fig:auto_encoder}(a,b), we test three training schemes for the QONN autoencoder.
(1) local-structured training [Fig.~\ref{fig:auto_encoder}(a), blue]: sequentially optimizing 2-layer QONNs to disentangle a single qubit at each stage, where each subsequent stage acts only on a reduced qubit subspace.  This approach is followed by a final global refinement step after all layers have been individually trained.
(2) global-structured training [Fig.~\ref{fig:auto_encoder}(a), orange]: where the above layer structure is trained simultaneously rather than sequentially. 
(3) global-unstructured training [Fig.~\ref{fig:auto_encoder}(b), green]: where a 6-layer system acting on all four qubits is trained. 

The optimization was performed using an implementation of \texttt{MLSL} \cite{Kan:1987ko} (also available in the \texttt{NLopt} library) which is a global optimization algorithm that explores the cost function landscape with a sequence of local optimizations (in this case \texttt{BOBYQA}) from carefully chosen starting points, using a heuristic to avoid local optima that have already been found.  
Our training states are the set of four ground-states of H$_2$ corresponding to bond lengths of 0.5, 1.0, 1.5, and 2.0 angstroms.
Both structured optimizations performed comparably, converging to a fidelity of 92.0\%.  However, we note that the the iterative approach could potentially be made more efficient if more stringent convergence criteria were introduced.  The unstructured optimization achieved a lower fidelity of 57.9\%.  It is unclear from our data whether the iterative approach would have better scaling or accuracy than the global optimization in an asymptotic setting.

% subsection quantum_optical_autoencoder (end)

\section{Quantum Reinforcement Learning}
Finally, to demonstrate the utility of QONNs for classical machine learning tasks, and to show that they continue to generalize in that setting, we examine a standard reinforcement learning problem: that of trying to balance an inverted pendulum \cite{barto1983neuronlike}. Classical deep reinforcement learning uses a policy network, i.e. a network that takes an observation vector as input and outputs a probability distribution over the space of allowed actions. This probability vector is then sampled to choose an action, a new observation is taken, and the process repeats. As the output from a QONN is inherently a probability distribution, policy networks are a natural application. 

\begin{figure}[t!]
  \begin{centering}
    \includegraphics[width=3.5in]{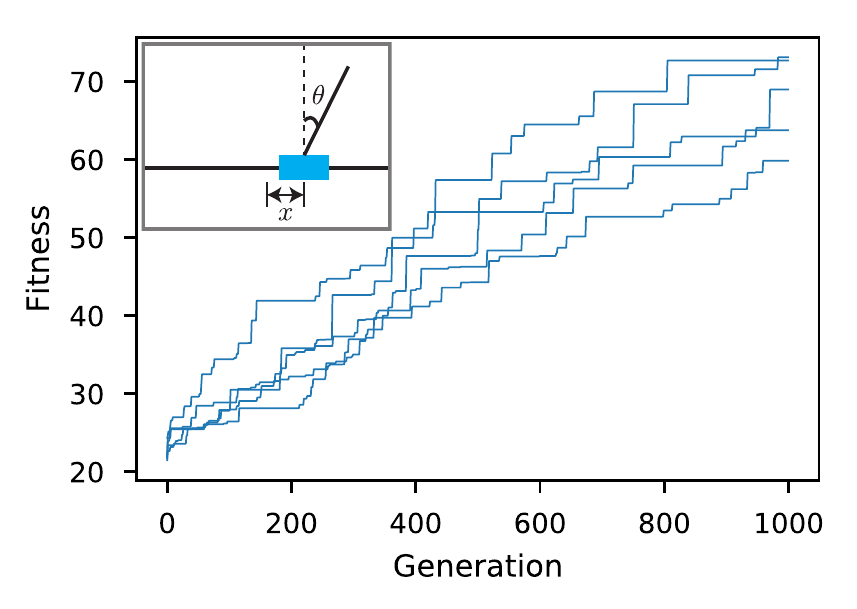}
  \end{centering}
  \caption{\label{fig:Cartpole} 
\textbf{Quantum Reinforcement Learning.} Fitness vs. training generation curves for five different training runs of the reinforcement learning QONN. A higher fitness corresponds to a network that was able to keep the pole upright and the cart within the bounds for more time. Input data to the QONN was encoded onto four qubits. Inset: The problem we are trying to solve, a cart on a bounded one-dimensional track with an inverted pendulum attached to the top. 
}
\vspace{-0.5cm}
\end{figure}

We simulate a cart moving on a one dimensional frictionless track, with a pole on a hinge attached to its top (see Fig.~\ref{fig:Cartpole}. inset). At the beginning of the simulation, the cart is initialized to a random position, with the pole at a random angle. At each timestep, the neural network receives four values, the position of the cart $x$, its velocity $\dot{x}$, the angle of the pole with respect to the track $\theta$, and the time derivative of that angle $\dot{\theta}$. From those four values, it determines whether to apply a force of unit magnitude either in the $+x$ or $-x$ directions; those are the only two options. Each run of the simulation continues until a boundary condition in $x$, $\theta$, or $t$ ($t_{max}=300$) is reached (i.e. the cart runs into the edge of the track or the pole falls over). The number of time steps before failure is the fitness of that run; we want to make this as large as possible.

To train a QONN to perform this task, we first encode the four values $x$, $\dot{x}$, $\theta$, and $\dot{\theta}$ onto four qubits. We do this by compressing the values for these each into the range $\gamma\in[0,\pi/2]$ and setting the respective input qubit to $\cos(\gamma)\ket{0} + \sin(\gamma)\ket{1}$. We then pass these four photons through a QONN and, at the output, use the first two modes to select an action. If the number of photons in the first mode exceeds the number in the second mode, we apply a force in the $-x$ direction; otherwise we apply a force in the $+x$ direction. Finally, we train these networks using an evolutionary strategies method \cite{salimans2017evolution}.

In Fig.~\ref{fig:Cartpole} we show the results of five training cycles (each with different starting conditions) using a 6-layer QONN. For each cycle, we use a batch size of 100 to determine the approximate gradient, and average the fitness over 80 distinct runs of the network at each $\vec{\Theta}$ we evaluate. Hyperparameters (layer depth, batch size, and averaging group) were tuned using linear sweeps. 
Fitness increases with training generation, meaning the QONN consistently learns to balance the pole for longer as time increases: generalizing examples it has previously seen to new instances of the problem.

To cross-check our performance we trained equivalently sized classical networks, i.e. 4-neuron, 6-layer networks with constant width. Hidden layers had ReLu neurons while the final layer was a single sigmoid neuron to generate a probability $p\in(0,1)$ of applying force in the $-x$ direction. 
We used the same training strategy for the classical networks as for the QONNs and observed a comparable performance, with a mean fitness after 1000 generations in the classical case of $37.1$ compared with $66.4$ for the QONN. 
Both networks can likely be optimized, and one should be cautious in directly comparing the classical and quantum results. 
Notwithstanding, this exploratory work demonstrates that quantum systems can learn on physically relevant data, and future directions will seek to leverage uniquely quantum properties such as superposition for batch learning \cite{Riste:2017ga}.

\section{Discussion}
%what we did
We have proposed an architecture for near-term quantum optical systems that maps many of the auspicious features of classical neural networks onto the quantum domain.
Through numerical simulation and analysis we have applied our QONN to a broad range of quantum information processing tasks, including newly developed protocols such as quantum optical state compression for quantum networking and black-box quantum simulation.
Experimentally, advances in integrated photonics and nano-fabrication have enabled monolithically integrated circuits with many thousands of optoelectronic components \cite{Chung:2017gj}, and a feasible route towards large-scale single photon readout \cite{DiZhu:2018ja}.
The architecture we present is not limited to the integration of systems with strong single photon nonlinearities, although promising progress has been made towards solid-state waveguide-based nonlinearities \cite{llner:2015kh, Sun57}.  
Rather, we anticipate our approach will serve as a natural intermediate step towards large-scale photonic quantum technologies.
In this intermediate regime, the QONN may learn practical quantum operations with weak or noisy nonlinearities which are otherwise unsuitable for fault-tolerant quantum computing \cite{Shapiro:2006fz}.
Future work will likely focus on loss correction techniques, which are also possible in an all optical context \cite{Niu:2018cq}.
Together, our results point towards both a powerful simulation tool for the design of next generation quantum optical systems, and a versatile experimental platform for near-term optical quantum information processing and machine learning.

\begin{acknowledgments}
This work was supported by the AFOSR MURI for Optimal Measurements for Scalable Quantum Technologies (FA9550-14-1-0052) and by the AFOSR program FA9550-16-1-0391, supervised by Gernot Pomrenke.
G.R.S. acknowledges support from the Facebook Fellowship Program.
J.C. is supported by EU H2020 Marie Sklodowska-Curie grant number 751016.
We gratefully acknowledge the support of NVIDIA Corporation with the donation of the Tesla K40 GPU used for this research.
\end{acknowledgments}

\appendix

\section{Computational Techniques} % (fold)
\label{sub:training}
The quantum optics simulations in this work were performed with custom, optimized code written in Python, with performance-sensitive sections translated to Cython. The Numba library was used to GPU accelerate some large operations. The most computationally intensive step was the calculation of the multi-photon unitary transform ($U(\vec{\theta}_i)$ in Eq.~\ref{eq1}) from the single photon unitary, which involves the calculation of the permanent of ${n+m-1 \choose n}^2$ matrices of dimension $n \times n$ \cite{Scheel:2004tt}. 
%We plan to open-source the optimized code written for this research after publication to ease future quantum optics simulation.  

As with classical neural networks, different optimization algorithms perform better for different tasks. We rely on gradient-free optimization techniques, as computing and backpropagating the gradient through the system likely requires knowledge of the internal quantum state of the system, preventing efficient training. While this might be acceptable for designing small systems in simulation (say, designing quantum gates), it doesn't allow for systems to be variationally trained in situ.  We empirically determined that the \texttt{BOBYQA} algorithm \cite{Powell:2gDjtIQ0} performs well for most applications in terms of speed and accuracy for our QONN, and is available in the \texttt{NLopt} library \cite{Johnson:2011}. For the quantum reinforcement learning experiments, we used our own implementation of evolutionary strategies \cite{salimans2017evolution}.

The computer used to perform these simulations is a custom-built workstation with a 12-core Intel Core i7-5820K and 64GB of RAM. The GPU used was an Nvidia Tesla K40. Relevant software versions are: Ubuntu 16.04 LTS, Linux 4.13.0-39-generic \#44~16.04.1-Ubuntu SMP, Python 2.7.12, NumPy 1.14.1, NLopt 2.4.2, Cython 0.27.3, and Numba 0.37.0. 

% subsection training (end)

\section{Benchmarking Training} % (fold)
\label{sub:be}
The training set for the Bell-state projector is the full set of Bell states $ \{ \ket{\psi^i_\text{in}} \} = \{\ket{\Phi^{+}}, \ket{\Phi^{-}}, \ket{\Psi^{+}}, \ket{\Psi^{-}}\}$ encoded as dual rail qubits. Our goal is to map these to a set of states distinguishable by single photon detectors thus we opt for a binary encoding $\{\ket{\psi^i_\text{out}}\}=\{ \ket{1010}, \ket{1001}, \ket{0110}, \ket{0101}  \}$.
A system designed to perform this map can then be run in reverse to generate Bell states from input Fock states.
The CNOT gate uses a full input-output basis set with $\{ \ket{\psi^i_\text{in}} \} = \{\ket{1010}, \ket{1001}, \ket{0110}, \ket{0101}\}$ and $\{ \ket{\psi^i_\text{out}} \} = \{\ket{1010}, \ket{1001}, \ket{0101}, \ket{0110}\}$. 
For the GHZ generator we select just a single input-output configuration $\{ \ket{\psi^i_\text{in}} \} = \{\ket{101010}\}$ and $\{ \ket{\psi^i_\text{out}} \} = \{(\ket{101010}+\ket{010101})/\sqrt{2} \}$. 
% subsection be (end)

\section{Simulated Hamiltonians} % (fold)
\label{sub:simulated_hamiltonians}
The Ising model we simulate is described by the Hamiltonian
\begin{equation}
	H_\text{ising} = B \sum_i \hat{X}_i + J \sum_{\langle i,j \rangle} \hat{Z}_i \otimes \hat{Z}_j ,
\end{equation}
where $B$ represents the interaction of each spin with a magnetic field in the $x$ direction, and $J$ is the interaction strength between spins in an orthogonal direction.
The Bose-Hubbard model we simulate is described by the Hamiltonian
\begin{equation}
	\hat{H}_\text{BH}= \omega \sum_i  \hat{b}^\dagger_i \hat{b}_i - t_\text{hop}\sum_{\langle i,j \rangle} \hat{b}^\dagger_i \hat{b}_j + U/2 \sum_i \hat{n}_i (\hat{n}_i-1),
\end{equation}  
where $\hat{b}_i^\dagger$ ($\hat{b}_i$) represents the creation (annihilation) operator in mode $i$, $\hat{n}_i$ the number operator and $\omega$, $t_\text{hop}$ and $U$ the on-site potential, the hopping amplitude and the on-site interaction strength respectively.

% subsection simulated_hamiltonians (end)

% section methods (end)

%

\end{document}